\documentstyle[12pt]{article}
\begin{document}

\title{Burgers' Flows as Markovian Diffusion Processes}
 
\author{Piotr Garbaczewski, Grzegorz Kondrat, Robert Olkiewicz\\ 
Institute of Theoretical Physics, University of Wroc{\l}aw,
pl. M. Borna 9, \\
PL-50 204 Wroc{\l}aw, Poland}

\maketitle
\hspace*{1cm}

\begin{abstract}

We analyze  the  unforced and deterministically forced  Burgers
equation in the framework  of   the
(diffusive) interpolating dynamics that solves  the so-called
Schr\"{o}dinger  boundary data problem for the
random matter transport.
This entails an exploration of the consistency conditions
that allow to interpret dispersion of passive contaminants in the
Burgers flow as a Markovian diffusion process.
In general, the
usage of a continuity equation 
$\partial _t\rho =-\nabla (\vec{v}\rho )$,
where  $\vec{v}=\vec{v}(\vec{x},t)$ stands for the Burgers field and 
$\rho $ is the  density of transported matter, is at variance with the 
explicit diffusion scenario. 
Under these circumstances, we give a complete 
characterisation of the  diffusive transport that is
governed by  Burgers velocity fields.  The result extends both to
the approximate description of the transport driven by an 
incompressible   fluid   and to motions in an infinitely 
compressible medium. Also, in conjunction with the Born statistical
postulate in quantum theory, it pertains to the probabilistic
(diffusive) counterpart of the Schr\"{o}dinger picture quantum
dynamics.\\
We give a generalisation of this dynamical problem to cases governed
by nonconservative  force fields when it appears indispensable  to
relax the gradient velocity field assumption.
The Hopf-Cole procedure has been appropriately
generalised to yield solutions in that case.

\end{abstract}

\newpage

\section{Burgers velocity fields and the related  stochastic
transport processes}

The Burgers equation, \cite{burg,hopf}, recently has acquired a
considerable popularity
in the variety of physical contexts, \cite{zeld}-\cite{walsh}.
  An exhaustive discussion of its role
in accoustic turbulence and gravitational contexts, where the
emergence
of shock pressure fronts is crucial, can be  found in the monograph
\cite{gurba}.\\
As is well known, the logarithmic Hopf-Cole transformation,
\cite{hopf}, allows  to replace  the nonlinear problem
("nonlinear diffusion equation", \cite{burg})  by a linear parabolic
equation. Because of this equivalence all gradient-type solutions
of the Burgers equation are  known exactly.

At the moment  we shall preserve the gradient form restriction for
Burgers velocity fields,
but consider a more general form of the Burgers equation that
accounts for an external force field $\vec{F}(\vec{x},t)$:
$${\partial _t\vec{v} + (\vec{v}\cdot \nabla )\vec{v} = \nu
\triangle \vec{v} +
\vec{F}(\vec{x},t)\enspace .} \eqno (1)$$
Let us mention that many of recent investigations were devoted to
the analysis of
$curl\, \vec{v}=\vec{0}$ solutions that are statistically relevant in
view of the random initial data choice and/or inclusion of the
random forcing term (the random potential in
the  related Parisi-Kardar equation, \cite{parisi}). \\
However, irrespective of whether we need or need not the
statistical input, an issue of
matter transport driven by those nonlinear velocity fields needs the
knowledge of an exact evolution of concentration  and/or density
fields.
Much in the spirit of early hydrodynamical studies of advection and
diffusion of passive tracers \cite{saf,town}, see also \cite{sai}. \\
This particular issue is addressed in the present paper, under a
simplifying assumption of nonrandom initial data and deterministic
force fields.\\
Following the traditional motivation (applicable both
to incompressible and infinitely compressible liquids, \cite{burg}),
we  regard the stochastic diffusion process as a primary phenomenon
responsible for the emergence of (1) and thus justifying the
"nonlinear diffusion equation" phrase in this context.

Knowing the  Burgers velocity  fields, one is tempted to   ask
what is the particular  dynamics (of matter or probability
density fields) that is consistent with the  chosen Burgers
velocity field evolution.
The corresponding passive scalar (tracer or contaminant)
advection-in-a-flow  problem, \cite{siggia,parisi,monin}, is
normally introduced through the  parabolic dynamics:
$${\partial _tT + (\vec{v}\cdot \nabla )T = \nu \triangle T\, }
\eqno (2)$$
see e. g.  also \cite{saf,town,sai}. For incompressible fluids, (2)
coincides with the conventional Fokker-Planck equation for the
diffusion process. This feature
does not persist in the compressible  case. \\
While looking for the stochastic implementation of the microscopic 
(molecular) dynamics (2), \cite{woy2,parisi,monin,sai},
it is  assumed
that the  "diffusing scalar" (contaminant  in the lore of early 
statistical turbulence models)  obeys an It\^{o} equation:  
$${d\vec{X}(t) =\vec{v}(\vec{x},t)dt  + \sqrt{2\nu } d\vec{W}(t)}
\eqno (3)$$
$$\vec{X}(0)=\vec{x}_0 \rightarrow \vec{X}(t)=\vec{x}$$
where the given  forced Burgers velocity field is    
perturbed by the noise term representing a molecular diffusion.
In the, by now conventional,  It\^{o} representation of diffusion-type
random variable
$\vec{X}(t)$ one explicitly refers to the standard Brownian motion
(e.g. the Wiener process) $\sqrt {2\nu }\vec{W}(t)$,
instead of the usually adopted formal white noise integral 
$\int_0^t \vec{\eta }(s)ds$, coming from the  Langevin-type version
of (3).

Under these premises, while taking for granted that \it there is \rm
a diffusion process involved, we cannot view equations (1)-(3) as
completely independent (disjoint) problems:
the velocity field $\vec{v}$ cannot be  quite arbitrarily
inferred from  (1) or any other velocity-defining equation without
verifying the \it consistency \rm conditions, which would allow to
associate (2) and (3)  with 
a well defined random dynamics, and
Markovian diffusion in particular, \cite{fried,horst}.

In connection with the usage of Burgers velocity fields (with
or without external forcing) which in (3) clearly are
intended to replace the 
standard \it forward  drift \rm of the would-be-involved Markov
diffusion process,  we have not found in the literature any 
attempt to resolve apparent contradictions arising if (2) and/or (3) 
are  defined by means of (1).
In particular, the
usage of a continuity equation 
$\partial _t\rho =-\nabla (\vec{v}\rho )$, 
where  $\vec{v}=\vec{v}(\vec{x},t)$ stands for the Burgers field and 
$\rho $ is the  density of transported matter, is at variance with the 
explicit diffusion scenario. 
Also, an issue of the necessary \it correlation \rm (cf.
\cite{monin},
Chap.7.3, devoted to the turbulent transport and the related
dispersion of contaminants)  between the probabilistic
Fokker-Planck
dynamics of the diffusing tracer, and this of the  passive tracer 
(contaminant)  concentration (2),  has been  left aside in the
literature. 

Moreover, rather obvious hesitation could have been  observed
in attempts to establish the  most appropriate  matter transport 
rule,  if (1)-(3)  are adopted. 
Depending  on the particular phenomenological departure point,
 one either adopts the standard continuity equation,
 \cite{zeld,alb}, that is certainly
valid to a high degree of  accuracy in the low viscosity limit
(we refer to the standard terminology that comes from viscous
fluid models; here, $\nu $ stands for the diffusion constant)
$\nu \downarrow 0$  of (1)-(3),  but incorrect on
mathematical grounds \it if \rm  there is a
diffusion involved \it and \rm simultaneously a solution of (1)
is interpreted as the respective \it current \rm velocity of the flow:
${\partial _t\rho (\vec{x},t)= - \nabla 
[\vec{v}(\vec{x},t)\rho (\vec{x},t)]\enspace . }
$ 
Alternatively, following  the white noise calculus tradition telling
that  the stochastic integral
$\vec{X}(t)=\int_{0}^{t} \vec{v}(\vec{X}(s),s)ds +
\int_{0}^{t} \vec{\eta }(s)ds$
 implies the Fokker-Planck equation, one adopts,
\cite{woy2}: ${\partial _t\rho (\vec{x},t) =
\nu \triangle \rho (\vec{x},t) -
\nabla [\vec{v}(\vec{x},t)\rho (\vec{x},t)]}$
which is clearly problematic in view of the classic Mc Kean's
discussion
of the  propagation of  chaos for the Burgers equation,
\cite{kean,cald,osa} and the derivation of the stochastic 
"Burgers process" in this  context: 
"the fun begins in trying to describe this Burgers motion as the 
path of a tagged molecule in an infinite bath of like molecules", 
\cite{kean}.

To put things on the solid ground, let us consider a Markovian
diffusion  process, which is characterised by the
transition probability density
(generally inhomogeneous in space and time law of random
displacements)
$p(\vec{y},s,\vec{x},t)\, ,\, 0\leq s<t\leq T$, and the probability 
density $\rho (\vec{x},t)$  of its random variable 
$\vec{X}(t)\, ,\,  0\leq t \leq T$. The process is completely 
determined  by these data. For clarity of discussion, we do not
impose  any spatial boundary  restrictions, nor fix any concrete 
limiting value of $T$ which, in principle, can be moved to infinity.

The conditions valid for any $\epsilon >0$: \\
(a) there holds
$lim_{t\downarrow s}{1\over
{t-s}}\int_{|\vec{y}-\vec{x}|>\epsilon }
p(\vec{y},s,\vec{x},t)d^3x=0$,\\
(b) there exists a (forward) drift
$\vec{b}(\vec{x},s)=lim_{t\downarrow s}{1\over {t-s}}\int_{|\vec{y}-
\vec{x}|
\leq \epsilon }(\vec{y}-\vec{x})p(\vec{x},s,\vec{y},t)d^3y$, \\
(c) there exists  a diffusion function (in our case  it is simply
a diffusion coefficient $\nu $)
$a(\vec{x},s)=lim_{t\downarrow s}{1\over {t-s}}
\int_{|\vec{y}-\vec{x}|\leq \epsilon }  (\vec{y}-\vec{x})^2
p(\vec{x},s,\vec{y},t)d^3y$,\\
are conventionally interpreted to define a diffusion process,
\cite{horst,fried}.
Under suitable restrictions  the function:
$${g(\vec{x},s)=
\int  p(\vec{x},s,\vec{y},T)g(\vec{y},T) d^3y }\eqno (4)$$
satisfies the backward diffusion equation (notice that the 
minus sign appears, in comparison  with (2))
$${- \partial _sg(\vec{x},s) = \nu \triangle g(\vec{x},s)  +
[\vec{b}(\vec{x},s)\cdot \nabla ]g(\vec{x},s)
\enspace .}\eqno (5)$$
Let us point out that the validity of (5) is known to be a \it
necessary
 \rm condition for the existence of a Markov diffusion process, whose
probability density $\rho (\vec{x},t)$ is to obey the Fokker-Planck 
equation. Here,  the new velocity field, named the forward drift
of the process $\vec{b}(\vec{x},t)$, replaces the
previously  utilized Burgers field $\vec{v}(\vec{x},t)$)): 
$${\partial _t\rho (\vec{x},t) = \nu \triangle 
\rho (\vec{x},t) - \nabla \cdot [\vec{b}(\vec{x},t)
\rho (\vec{x},t)]\enspace .}
\eqno (6)$$

The case  of particular interest in the  
nonequilibrium statistical physics literature  
appears when $p(\vec{y},s,\vec{x},t)$
is  a \it fundamental solution \rm of (5) with respect to variables
$\vec{y},s$,  \cite{krzyz,fried,horst},
see however \cite{olk2} for  an alternative situation.
Then, the transition probability density satisfies \it also \rm
 the Fokker-Planck equation in the
remaining $\vec{x}, t$ pair of variables.  Let us emphasize that
these two equations form an adjoint pair,
referring to the  slightly counterintuitive for
physicists, although transparent for
mathematicians, \cite{haus,fol,nel,zambr,zambr1}, issue of time
reversal of diffusion processes.

After  adjusting (3) to the present  context, $\vec{X}(t)=
\int_0^t\, \vec{b}(\vec{X}(s),s)\, ds + \sqrt {2\nu } \vec{W}(t) $
we realise,
\cite{nel1,nel,zambr,zambr1},  that for any smooth function
$f(\vec{x},t)$ of the random variable $\vec{X}(t)$ the 
conditional expectation value:
$$lim_{\Delta  t\downarrow 0} {1\over {\Delta t}}\bigl [\int
p(\vec{x},t,\vec{y},t+
\Delta
t)f(\vec{y},t+\Delta t)d^3y - f(\vec{x},t)\bigr ] =
(D_+f)(\vec{X}(t),t)=    \eqno (7) $$
$$=  (\partial _t + (\vec{b}\cdot \nabla )+ \nu \triangle )
f(\vec{x},t)$$
$$\vec{X}(t)=\vec{x}$$
determines  the forward drift $\vec{b}(\vec{x},t)$ (if we set 
components of $\vec{X}$ instead of $f$) and allows to introduce  
the local field of (forward) accelerations associated with the
diffusion process, which we constrain by demanding (see e.g. Refs. 
\cite{nel1,nel,zambr,zambr1} for prototypes of such dynamical 
constraints):
$${(D^2_+\vec{X})(t) =(D_+\vec{b})(\vec{X}(t),t) =(\partial _t
\vec{b} +
(\vec{b}\cdot \nabla )\vec{b} + \nu \triangle
\vec{b})(\vec{X}(t),t)= \vec{F}(\vec{X}(t),t)}\eqno (8)$$
where, at the moment arbitrary,  function $\vec{F}(\vec{x},t)$ 
may be interpreted as the external deterministic forcing applied to
the diffusing system,  \cite{blanch}. 
In particular, if we assume that drifts remain
gradient fields, $curl \, \vec{b}= \vec{0}$, under the forcing, then
those that are allowed by the prescribed choice of
$\vec{F}(\vec{x},t)$    \it must \rm fulfill the compatibility
condition (notice the conspicuous absence of the standard
Newtonian minus sign in this analogue of the second Newton law)
$${\vec{F}(\vec{x},t)=\nabla \Omega (\vec{x},t) }\eqno (9)$$
$$\Omega (\vec{x},t) = 2\nu \bigl [\partial _t \Phi \, +\,
{1\over 2} ({\vec{b}^2
\over {2\nu }}+ \nabla \cdot \vec{b})\bigr ] \enspace .$$
This establishes the  connection of the
forward drift
$\vec{b}(\vec{x},t)=2\nu \nabla \Phi (\vec{x},t)$ with the
(Feynman-Kac,
cf. \cite{blanch,olk2}) potential $\Omega (\vec{x},t)$ of the
 chosen external force field. The latter connection, without
 invoking the Feynman-Kac formula,  is frequently
 exploited in the theory of Smoluchowski-type diffusion processes,
 when the Fokker-Planck equation is transformed into the associated
 generalised diffusion  equation.

One of distinctive features of Markovian diffusion processes
with the positive density $\rho (\vec{x},t)$ is that the
notion of the \it backward \rm transition probability density
 $p_*(\vec{y},s,\vec{x},t)$ can be consistently introduced on 
each finite  time interval, say $0\leq s<t\leq T$:
 $${\rho (\vec{x},t) p_*(\vec{y},s,\vec{x},t)=p(\vec{y},s,\vec{x},t)
 \rho (\vec{y},s)} \eqno (10)$$
so that $\int \rho (\vec{y},s)p(\vec{y},s,\vec{x},t)d^3y=
\rho (\vec{x},t)$ 
and $\rho (\vec{y},s)=\int p_*(\vec{y},s,\vec{x},t)
\rho (\vec{x},t)d^3x$.  
This  allows to define (cf. \cite{nel1,blanch,vig,olk} for a
discussion of
these concepts in case of the most traditional Brownian motion and
Smoluchowski-type diffusion processes)
$$lim_{\Delta t\downarrow 0} \, {1\over {\Delta t}}\bigl
[ \vec{x} - \int p_* (\vec{y},t-\Delta t,\vec{x},t)\vec{y} d^3y
\bigr ]= (D_-\vec{X})(t)=
\vec{b}_*(\vec{X}(t),t)
\eqno (11) $$
$$(D_-f)(\vec{X}(t),t) = (\partial _t + (\vec{b}_* \cdot \nabla )-
\nu
\triangle )f(\vec{X}(t),t)$$
Accordingly, the backward version  of the dynamical constraint 
imposed on the local acceleration field  reads
$${(D^2_-\vec{X})(t) = (D^2_+\vec{X})(t) = \vec{F}(\vec{X}(t),t) }
\eqno (12) $$
where under the gradient-drift field assumption, $curl \, \vec{b}_*=0$ 
we have  explicitly involved the forced Burgers equation (cf. (1)):
$${\partial _t\vec{b}_* +  (\vec{b}_*\cdot \nabla )\vec{b}_* -
\nu \triangle \vec{b}_* 
= \vec{F}} \eqno (13)$$
Here, \cite{nel,zambr,blanch}, in view of
$\vec{b}_*= \vec{b} - 2\nu \nabla ln \rho $, we deal 
with $\vec{F}(\vec{x},t)$ previously introduced in  (9).
A notable consequence  is that the Fokker-Planck equation (6) can be
transformed to an \it equivalent \rm  form of:
$${\partial _t\rho(\vec{x},t) = - \nu \triangle \rho (\vec{x},t) - 
\nabla [\vec{b}_*(\vec{x},t) \rho (\vec{x},t)]}\eqno (14)$$
which however describes a density evolution in the reverse sense of
time (!).

At this point let us recall that  equations (5) and (6) form a natural 
 adjoint  pair of equations that determine the Markovian  
diffusion process in the chosen time interval $[0,T]$.  
Clearly, an adjoint of  (14), reads:
$${\partial _s f(\vec{x},s) = \nu \triangle f(\vec{x},s) - 
[\vec{b}_*(\vec{x},s)\cdot \nabla ] f(\vec{x},s)}\eqno (15)$$
where:
$${f(\vec{x},s)=\int p_*(\vec{y},0,\vec{x},s) f(\vec{y},0)d^3y
\enspace  , }\eqno (16)$$
to be compared with (4),(5) and the previously mentioned 
passive scalar dynamics (2), see e.g. also \cite{woy2}.
Here, manifestly, the time evolution of
the backward drift is governed by the Burgers equation, 
and the diffusion equation (15) is correlated (via the
definition (10)) with the probability density evolution rule (14). 

This pair \it only \rm can be consistently utilized if the 
diffusion proces  is to be  driven by forced (or unforced) 
Burgers velocity fields. Certainly, the continuity equation
postulated to involve the Burgers field as the current
velocity, does not hold true in this context.

Let us point out that the  study of diffusion in 
the Burgers flow may begin from first solving the Burgers equation
(12) for a chosen external force field, next 
specifying the  probability density 
evolution (14), eventually ending with the corresponding "passive 
contaminant" concentration dynamics (15), (16). All that remains in 
perfect agreement with  the heuristic discussion of the 
concentration dynamics given in   Ref. \cite{monin}, Chap. 7.3. 
where the "backward dispersion" problem 
with "time running backwards" was found necessary to \it predict \rm 
the concentration.

All that means that equations (1)-(3) can be reconciled in the
framework set by (4)-(16). Then,  the "nonlinear diffusion
equation" does indeed refer to consistent stochastic diffusion
processes.

We are now at the point where the Burgers equation
and the related matter  transport can be consistently embedded in
the general probabilistic framework of the so-called
Schr\"{o}dinger's boundary data (stochastic interpolation)
problem, \cite{zambr,zambr1,olk2,blanch,olk,olk1},
see also \cite{alb1,freid}. In this setting,
the familiar Hopf-Cole transformation,
\cite{hopf,flem}, of the Burgers equation into the 
generalised diffusion equation (yielding explicit solutions in the 
unforced case) receives a useful generalisation.

Indeed, in that framework, \cite{blanch,olk2}, the problem of
deducing  a suitable  Markovian diffusion process  was reduced to
investigating the  adjoint pairs  of parabolic partial 
differential equations, like e.g. (5), (6) or (14), (15). 
In case of gradient drift fields  this amounts to  
checking  (this imposes limitations on the admissible 
force field potential, cf. also
 formula (9)) whether  the Feynman-Kac kernel
$${k(\vec{y},s,\vec{x},t)=\int exp[-\int_s^tc(
\omega (\tau ),\tau)d\tau ]
d\mu ^{(y,s)}_{(x,t)}(\omega )}\eqno (17) $$
is positive and continuous in the open space-time
area of interest, and whether it gives rise to positive solutions of
the adjoint pair of  generalised heat equations:
$${\partial _tu(\vec{x},t)=\nu \triangle u(\vec{x},t) -
c(\vec{x},t)u(\vec{x},t)}\eqno (18)$$
$$\partial _tv(\vec{x},t)= -\nu \triangle v(\vec{x},t) +
c(\vec{x},t)v(\vec{x},t)$$
where  $c(\vec{x},t)={1\over {2\nu }} \Omega (\vec{x},t)$ follows
from the previous formulas.
In the above, $d\mu ^{(\vec{y},s)}_{(\vec{x},t)}(\omega)$ is the
conditional
Wiener measure over sample paths of the standard Brownian motion.

Solutions of (18), upon suitable normalisation give rise to the
Markovian  diffusion process with the factorised probability density
$\rho (\vec{x},t)=u(\vec{x},t)v(\vec{x},t)$ wich interpolates between 
the boundary density data $\rho (\vec{x},0)$ and 
$\rho (\vec{x},T)$, with the forward and backward 
drifts of the process defined as follows:
$${\vec{b}(\vec{x},t)=2\nu {{\nabla v(\vec{x},t)}
\over {v(\vec{x},t)}}}
\eqno (19)$$
$$\vec{b}_*(\vec{x},t)= - 2\nu {{\nabla u(\vec{x},t)}
\over {u(\vec{x},t)}}$$
in the prescribed time interval $[0,T]$. 
The transition probability density of this process reads:
$${p(\vec{y},s,\vec{x},t)=k(\vec{y},s,\vec{x},t)
{{v(\vec{x},t)}\over {v(\vec{y},s)}}\enspace ,}\eqno (20)$$
Here, neither  $k$, (17), nor $p$, (20) need to be the 
fundamental  solutions  of appropriate parabolic equations, 
see e.g. ref. \cite{olk2} where an issue of  
differentiability is analyzed.

The corresponding (since $\rho (\vec{x},t)$ is given) transition
probability density, (10),  of the backward process has the form:
$${p_*(\vec{y},s,\vec{x},t) = k(\vec{y},s,\vec{x},t)
{{u(\vec{y},s)}\over {u(\vec{x},t)}}\enspace .}\eqno (21)$$
Obviously, \cite{olk2,zambr}, in the time interval $0\leq s<t\leq T$
there holds: \\
${u(\vec{x},t)=\int u_0(\vec{y}) k(\vec{y},s,\vec{x},t) d^3y}
$ and
$v(\vec{y},s)=\int k(\vec{y},s,\vec{x},T) v_T(\vec{x})d^3x$.

By defining $\Phi _*=log\, u$, 
we immediately recover the traditional form of the Hopf-Cole
transformation 
 for Burgers velocity fields: $\vec{b}_*=-2\nu \nabla \Phi _*$.
In the special case of the standard free Brownian motion, there 
holds $\vec{b}(\vec{x},t)=\vec{0}$ while $\vec{b}_*(\vec{x},t)=
-2\nu
\nabla log\, \rho (\vec{x},t)$. 

Our discussion provides  a complete identification of the
 stochastic diffusion process underlying \it both \rm the
 deterministically forced Burgers
 velocity dynamics and the related matter transport, (14), the latter
 in terms of suitable density fields.
 The generalisation of the Hopf-Cole procedure
 to this case involves a powerful methodology of Feynman-Kac  kernel
 functions and yields exact formulas for solutions for the forced
Burgers equation.
Let us stress that the connection between the Burgers
 equation and the generalised (forward) heat  equation is not
 merely a
 formal trick that generates solutions to the nonlinear problem.
 The forward equation (18) in fact carries a complete
 information about
 the implicit \it backward stochastic evolution \rm , that is a
 Markov diffusion process for which the Burgers-velocity driven
 transport is  appropriate.
 Notice that  the transition probability density (21) obeys the
 familiar
 Chapman-Kolmogorov formula. If we wish to analyze a concrete density
 field governed by this process, any two boundary density data $\rho
 (\vec{x},0)$ and $\rho (\vec{x},T)$ allow to deduce the ultimate form
 of the (more traditional, forward) diffusion process (20), by means of
 the Schr\"{o}dinger boundary data problem, \cite{zambr,olk2}. Then,
 the adjoint pair of equations (18) gives all details of the dynamics,
 with (19)-(21) as a necessary consequence. On the other hand, the
 presented discussion implies a direct import of the shock-type matter
 density profiles to the general nonequilibrium statistical physics of
 diffusion-type processes.

\section{The problem of nonconservative  forcing of Burgers velocity
fields}

By embedding the Burgers equation in the Schr\"{o}dinger's
interpolation  framework, we could consistently handle random
transport that is governed by gradient velocity fields and
gradient-type external conservative forces.
The natural question at this point is how to incorporate the
 the non-gradient (rotational for
example) velocity fields  and especially  the nonconservative
forces. This  question may be addressed without reservations only
in the context of the
forced Burgers  equation.
Recall that  the Hopf-Cole transformation is applicable
only in case of gradient velocity fields.
Moreover, the involved  Schr\"{o}dinger's interpolation framework
extends the
issue to  the domain of nonequilibrium random phenomena,
where standard
Smoluchowski diffusions, \cite{blanch},  are normally discussed
in the case of conservative force fields
(and drifts in consequence).\\

{\bf Remark}:
Strikingly, an investigation of typical nonconservative e.g.
electromagnetically forced diffusions
has not been much pursued in the literature, although an issue of 
deriving the Smoluchowski-Kramers equation (and possibly its large 
friction limit) from the Langevin-type equation for the charged 
Brownian particle in the general electromagnetic field  has 
been relegated in Ref. \cite{schuss}, Chap. 6.1 to the status of 
the innocent-looking exercise (sic !).  
On the other hand, the diffusion
of realistic charges in dilute ionic solutions creates a number of 
additional difficulties due to the apparent Hall mobility in terms 
of mean currents induced by the electric field (once assumed to act 
upon the system), see e.g. \cite{hub,sung} and \cite{mori}.
In connection with the electromagnetic forcing of diffusing charges,
the gradient field
assumption imposes  a severe limitation if we account for typical
(nonzero circulation) features of the classical
motion  due to the Lorentz force, with or  without the random
perturbation component.  
The purely electric forcing is simpler to handle, since it has 
a definite gradient field  realisation, see e.g. \cite{izm} for a
recent discussion of related issues.
The major obstacle with respect to our previous (Section 1)
discussion is that, if we wish to regard either
the force  $\vec{F}$, (8), (12),
or  drifts $\vec{b}$, $\vec{b}_*$  to have an electromagnetic
origin, then necessarily we need to pass from conservative
to non-conservative  fields. This subject matter has not been
significantly exploited so far in the nonequilibrium statistical
physics literature. \\

With this additional (to Burgersian per se) motivation, let us
analyse how the gradient velocity field (and  conservative force
field) assumption can be relaxed and  nonetheless the
exact  solutions to the Burgers equation  can be obtained,
\it both \rm in the unforced and forced cases, while involving the
primoridal Markovian diffusion process scenario. \\

It turns out the the crucial point of our previous discussion 
lies in a \it proper \rm choice of the strictly positive and
continuous, in an open space-time area,  
function $k(\vec{y},s,\vec{x},t)$ which, if we wish to 
construct a Markov  process, 
has to satisfy the Chapman-Kolmogorov (semigroup composition) 
equation.  It has led us to 
consider a pair of adjoint partial differential equations, (18),
as an alternative to either (5), (6) or (14), (15).

The Feynman-Kac integration is predominantly utilised
in the quantally oriented literature dealing with Schr\"{o}dinger
operators and their spectral properties, \cite{simon,glimm}.
We shall exploit some of results of this well developed theory.
The pertinent Feynman-Kac potential $c(x,t)$ in (17), (18)
is usually assumed  to be a  continuous and bounded from below
function, but these restrictions can be substantially relaxed
(unbounded functions are allowed in principle) if we wish to
consider  general Markovian diffusion processes  and disregard
an issue of  the bound state spectrum and this of the ground state
 of the  (selfadjoint) semigroup generator, \cite{krzyz,fried}.
Actually, what we need is merely that properties of $c(\vec{x},t)$
allow for the kernel $k$, (17), that  is positive and continuous.
This property is crucial for the Schr\"{o}dinger boundary-data
problem analysis. \\
Taking for granted that suitable
conditions are fulfilled, \cite{simon,olk2},
we can immediately associate with  equations (18) an
integral kernel of the time-dependent semigroup
(the exponential operator should be understood as  time-ordered
expression, since in general $H(\tau )$ may not commute with
$H(\tau ')$ for $\tau \neq \tau '$):
$${k(\vec{y},s,\vec{x},t)=
[exp(-\int_s^t H(\tau )d\tau )](\vec{y},\vec{x})}\eqno (23)$$
where $H(\tau )=-\nu \triangle +c(\tau )$ is the pertinent
semigroup generator.
Then, by the Feynman-Kac formula, \cite{freid}, we get an
expression (17) for the kernel, which in turn yields (19)-(22),
see e.g. \cite{olk2}.
As mentioned before, (20) combined with (17) sets a
probabilistic connection between the Wiener measure
corresponding to the standard Brownian motion with
$\vec{b}(\vec{x},t)=\vec{0}$, and this appropriate
for the diffusion process with a nonvanishing  drift
$\vec b(\vec{x},t),\, curl \, \vec{b} =\vec{0}$.

Our main purpose is to generalise  (23), so that the positive
and continuous (semigroup) kernel function can be associated with
stochastic diffusion processes,
whose drifts are no longer gradient-fields. In particular, the forcing
is to be nonconservative.\\

Since we have no particular hints towards  Feynman-Kac type analysis
of rotational motions,  it seems instructive to invoke the
framework of the Onsager-Machlup approach
towards an identification of most probable paths
associated with the underlying diffusion process,
\cite{has1,hunt,ronc}. In this context, the non-conservative model
system has been investigated in Ref. \cite{wieg}.
Namely, an  effectively  two-dimensional Brownian motion was
analyzed, whose three-dimensional forward
drift $\vec{b}(\vec{x}), \, b_3=0$ in view of
$\partial_xb_1\neq \partial _yb_2$
has $curl \, \vec{b}\neq 0$.
Then, by the standard variational
argument with respect to the Wiener-Onsager-Machlup action,
\cite{hunt,wieg},
$${I\{ L(\dot{\vec{x}},\vec{x},t);t_1,t_2\} = {1\over {2\nu }}
\int_{t_1}^{t_2} \{ {1\over 2}[\dot{\vec{x}} -
\vec{b}(\vec{x},t)]^2 +
\nu \nabla \vec{b}(\vec{x},t)\} dt \enspace ,}\eqno (24)$$
 the most probable trajectory, about which  major contributions
from  (weighted)  Brownian paths are concentrated,
was found to be a solution of the Euler-Lagrange equations,
which are formally identical to the equations of motion
$${\ddot{\vec{q}}_{cl}= \vec{E} +
\dot{\vec{q}}_{cl} \times \vec{B}}\eqno (25)$$
of a classical particle of unit mass and
unit charge moving in an electric field $\vec{E}$ and the magnetic
field $\vec{B}$.
The electric field (to be compared with Eq. (9)) is given by:
$${\vec{E}=-\nabla \Phi }\eqno (26)$$
$$\Phi = -{1\over 2} (\vec{b}^2 + 2\nu \nabla \vec{b})$$
while the magnetic field has the only nonvanishing component in
 the z-direction of $R^3$:
$${\vec{B}=curl \, \vec{b}= \{ 0,0,\partial _xb_2-\partial _yb_1\}
\enspace , } \eqno (27)$$
 Clearly, $\vec{B}= curl \vec{A}$ where
$\vec{A}\dot{=}\vec{b}$ is the electromagnetic vector potential.
The simplest  example  is a notorious constant magnetic
field defined by  $b_1(\vec{x})=-{B\over 2} x_2,\, b_2(\vec{x})=
{B\over 2} x_1$.

One immediately realizes that the Fokker-Planck equation in
this case is incompatible with  traditional intuitions
underlying the Smoluchowski-drift identification: the forward
drift is \it not \rm proportional to  an external force, but to an
electromagnetic potential.
Nevertheless, the variational
information  drawn from the Onsager-Machlup Lagrangian involves
the Lorentz force-driven trajectory.
Hence, some principal effects of the
electromagnetic forcing are present in the diffusing system, whose
drifts display an "unphysical" (gauge dependent) form.

On the other hand,  if we accept this "unphysical" random motion
to yield the representation  with the nongradient
drift $\vec{A}$:
$d\vec{X}(t)=\vec{A}(\vec{X}(t),t)dt +\sqrt{2\nu } d\vec{W}(t)$,
and consider
the corresponding pair (5), (6) of adjoint diffusion equations
with $\vec{A}(\vec{x},t)$  replacing $\vec{b}(\vec{x},t)$,
then  (8) tells us that
$${(D^2_+\vec{X})(t)=\partial _t\vec{A} + (\vec{A}\cdot \nabla )
\vec{A} +
\nu \triangle \vec{A} = - {{B^2}\over 4} \{ x_1,x_2,0 \} = -
\vec{E}(\vec{x})}\eqno (28)$$
where $\vec{E}(\vec{x})={{B^2}\over 4}\{ x_1,x_2,0 \} $, if
calculated from (26).

We thus arrive at the purely electric forcing with
reversed sign (if compared with that coming from the
Onsager-Machlup argument, (26)) and somewhat surprisingly,
there is no   impact of the previously discussed magnetic motion
 on the level of dynamical  constraints  (8), (13).
The adopted recipe is thus incapable of producing the magnetically
forced  diffusion process  that conforms with arguments of
Section 1.  Our toy model is inappropriate and more sophisticated
route must be  adopted.\\

In below, we shall invoke the Feynman-Kac kernel idea (23),
\cite{olk2}.
This approach has a clear  advantage of elucidating the
generic issues that hamper attempts to describe  the diffussion
processes  governed by nonconservative (and
electromagnetic in particular)  force fields.\\
The Burgers equation
and the problem of its nongradient solutions will appear residually as
a byproduct of the more general discussion.\\

Usually,  the selfadjoint semigroup generators attract the
attention of physicists in connection with the Feynman-Kac
formula. Since electromagnetic fields provide  most conventional
examples of nonconservative forces, we shall concentrate on their
impact on random dynamics. \\
 A typical route towards
incorporating electromagnetism comes from quantal motivations
via the minimal electromagnetic coupling recipe which
preserves the selfadjointness of the  generator (Hamiltonian
of the system).  As such, it constitutes a part of the general
theory of Schr\"{o}dinger operators.
A rigorous study of operators of the form
$-\triangle +V$ has become a well developed mathematical
discipline, \cite{simon}.
The study of Schr\"{o}dinger operators with magnetic fields,
typically of the form $-(\nabla -i\vec{A})^2 +V$,  is
less advanced, although specialised chapters on the magnetic field
issue can be found in monographs devoted to  functional
integration methods, \cite{simon,roep}, mostly in  reference  to
 seminal papers \cite{simon1,simon2}.

From the mathematical  point of view, it is desirable to deal with
 magnetic fields that go to zero at infinity, which is
certainly acceptable on physical grounds as well.  The
constant magnetic field (see e.g. our previous considerations)
does not meet this requirement, and its notorious usage in the
literature makes us (at the moment) to decline the asymptotic
assumption and inevitably fall into a number of serious
complications.

One obvious  obstacle can be seen immediately by taking
advantage of the
existing results, \cite{simon1}.  Namely, an explicit expression
for  the Feynman-Kac kernel in a constant magnetic  field,
introduced through the  the minimal electromagnetic coupling
assumption $H(\vec{A})=-{1\over 2}(\nabla -i\vec{A})^2$, is
available (up to irrelevant dimensional constants):
$${exp[- t H(\vec{A})](\vec{x},\vec{y})=
[{B\over {4\pi sinh({1\over 2}Bt)}}] ({1\over {2\pi t}})^{1/2}}
\eqno (29)$$
$$\times exp\{ - {1\over {2t}}(x_3-y_3)^2 - {B\over 4}
coth({B\over 2}t) [(x_2-y_2)^2+ (x_1-y_1)^2] -
i {B\over 2}(x_1y_2-x_2y_1)\} \enspace. $$
Clearly, it is  \it not  \rm   real (hence \it non-positive
\rm  and directly at variance with  the  major demand in the
Schr\"{o}dinger interpolation problem, as outlined in Section 1),
except for directions $\vec{y}$ that  are parallel to  a chosen
$\vec{x}$.

Consequently, a bulk of the well developed mathematical theory is
of no use for our purposes  and new techniques must be developed
for  a consistent description  of the electromagnetically
forced diffusion processes along the lines of Section 1, i.e.
within the framework of Schr\"{o}dinger's interpolation problem.
Also, another approach is necessary to generate solutions
of the Burgers equation that are not in the gradient form.

\section{Forcing via Feynman-Kac semigroups}

The conditional Wiener measure
$d\mu ^{(\vec{y},s)}_{(\vec{x},t)}(\vec{\omega })$,
appearing  in the Feynman-Kac kernel definition (17), if unweighted
(set $c(\vec{\omega }(\tau ),\tau)=0$)   gives rise to the
familiar heat kernel. This, in turn, induces the Wiener measure
$P_W$ of the set of
all sample paths,which originate from $\vec{y}$ at time $s$ and
terminate (can
be located) in the Borel set $A \in R^3$ after time $t-s$:
$P_W[A] = \int_A d^3x \int
d\mu ^{(\vec{y},s)}_{(\vec{x},t)}(\vec{\omega })
=\int_A d\mu $
where, for simplicity of notation, the $(\vec{y},t-s)$ labels are
omitted and $\mu ^{(\vec{y},s)}_{(\vec{x},t)} $ stands
for the heat kernel.

Having defined an It\^{o} diffusion
$\vec{X}(t)=\int_0^t \vec{b}(\vec{x},u)du + \sqrt{2\nu }
\vec{W}(t)$
we are interested in the analogous
path measure:
$P_{\vec{X}}[A] =
\int_A dx\int d\mu _{(\vec{x},t)}^{(\vec{y},s)}(
\vec{\omega }_{\vec{X}}) = \int _A d\mu (\vec{X})$.

Under suitable (stochastic, \cite{blanch}) integrability
conditions imposed on the forward drift, we have granted
the  absolute continuity $P_X \ll P_W$ of measures,
which implies the existence of a
strictly positive Radon-Nikodym density.  Its  canonical
Cameron-Martin-Girsanov form, \cite{simon,blanch}, reads:
$${{{d\mu (\vec{X})}\over {d\mu }}(\vec{y},s,\vec{x},t) =
exp  {1\over {2\nu }}\bigl [\int_s^t
\vec{b}(\vec{X}(u),u)d\vec{X}(u) - {1\over 2}
\int_s^t [\vec{b}(\vec{X}(u),u)]^2
du\bigr ] \enspace . } \eqno (30) $$

If we assume that drifts are gradient fields, $curl \, \vec{b}=0$,
then  the It\^{o} formula allows to reduce,
otherwise troublesome, stochastic
integration in the exponent of (30), \cite{simon,roep},  to
ordinary Lebesgue integrals:
$${{1\over {2\nu }}\int_s^t \vec{b}(\vec{X}(u),u)d\vec{X}(u)
= \Phi  (\vec{X}(t),t) - \Phi (\vec{X}(s),s) -
\int_s^t du\, [\partial _t\Phi  + {1\over 2} \nabla \cdot
\vec{b} ]
(\vec{X}(u),u) \enspace .  } \eqno (31) $$
After inserting (31) to (30) and next integrating with respect to
 the conditional Wiener measure, on account of (9) we
 arrive at the standard   form of the Feynman-Kac kernel (17).
Notice that (31) establishes a probabilistic basis for logarithmic
transformations (19) of forward and backward drifts:
$b=2\nu \nabla log\, v=2\nu \nabla \Phi $,
$b_*=-2\nu \nabla log \, u=-2\nu \nabla \Phi _*$. The forward
version is commonly  used in connection with the
transformation of the Fokker-Planck equation  into the
generalised heat equation, \cite{garb,blanch}.
The backward
version  is the Hopf-Cole transformation, mentioned in
Section 1, used to map the Burgers equation into the very
same generalised heat equation as in the previous case,
\cite{hopf,alb1}.

 However, presently we are interested in non-conservative
 drift fields, $curl \, \vec{b} \neq 0$, and in that case the
 stochastic integral in (30) is the major source of computational
 difficulties, \cite{simon,roep,nel}, for nontrivial vector
 potential field configurations.  It explains the virtual
 absence of magnetically forced diffusion problems in the
 nonequilibrium statistical physics literature.

 At this point, some  steps of the analysis performed in
 Ref. \cite{abc} in the context of the "Euclidean quantum
 mechanics", cf. also \cite{zambr1}, are extremely useful.
 Let us emphasize that  electromagnetic fields we utilize,  are
 always meant to be ordinary Maxwell fields with \it no \rm
 Euclidean connotations
 (see e.g. Chap.9 of Ref. \cite{roep} for the Euclidean version of
 Maxwell theory).

Let us consider a gradient drift-field diffusion problem  according
to Section 1, with (17), (31)  involved  and thus an  adjoint pair
(18) of parabolic equations completely defining the Markovian
diffusion process. Furthermore, let $\vec{A}(\vec{x})$ be the
time-independent vector potential for the Maxwellian
magnetic field $\vec{B}= curl  \, \vec{A}$.
We pass from the gradient realisation of drifts to  the
new  one, generalizing (19), for  which  the following
decomposition
into the gradient and   nonconservative part is valid:
$${\vec{b}(\vec{x},t)= 2\nu \, \nabla log\, \Phi (\vec{x},t) -
\vec{A}(\vec{x})\enspace ,} \eqno (32)$$
We  denote  $\theta (\vec{x},t)\dot{=}
exp\, [\Phi (\vec{x},t)]$
and admit that (32)  is a forward drift of an It\^{o} diffusion
process with a stochastic differential
${ d\vec{X}(t)= [2\nu {{\nabla \theta }\over {\theta }} -
\vec{A}]dt +  \sqrt {2\nu } d\vec{W}(t)}$.
On purely formal grounds, we deal here with  an example of the
Cameron-Martin-Girsanov transformation of the forward drift of
a given Markovian diffusion process and we are entitled to ask for
a corresponding measure transformation, (30).

To this end, let us furthermore  \it assume \rm that
$\theta (\vec{x},t)=\theta $ solves a partial differential equation
$${\partial _t\theta  =
 - \nu [ \nabla   - {1\over {2\nu }}
\vec{A}(\vec{x})]^2\theta  + c(\vec{x},t)\theta }\eqno  (33) $$
with the notation $c(\vec{x},t)={1\over {2\nu }}\Omega (\vec{x},t)$
patterned after (9).
 Then, by using  the It\^{o} calculus
and (32), (33) on the way, see e.g. Ref. \cite{abc}, we can
rewrite (30) as follows:
$${{{d\mu (\vec{X})}\over {d\mu }}(\vec{y},s,\vec{x},t)=}\eqno (34)$$
$${= exp\, {1\over {2\nu }} \bigl [
\int_s^t\bigl ( 2\nu {{\nabla \theta }\over {\theta }} -
\vec{A} \bigr ) (\vec{X}(u),u) d\vec{X}(u) - {1\over 2}
\int_s^t \bigl ( 2\nu {{\nabla \theta }\over {\theta }} -
\vec{A} \bigr )^2(\vec{X}(u),u)\, du\bigr ] }$$
$${=\, {{\theta (\vec{X}(t),t)}\over {\theta (\vec{X}(s),s)}}
\, exp\bigl [ - {1\over {2\nu }}  \int_s^t  [
\vec{A}(u)d\vec{X}(u) \, + \, \nu (\nabla \cdot
\vec{A})(\vec{X}(u))du \,
+\, \Omega (\vec{X}(u),u)du ] \bigr ]\enspace , } $$
where $\vec{X}(s)=\vec{y}, \vec{X}(t)=\vec{x}$.

More significant observation is that  the Radon-Nikodym density (34),
if integrated  with respect to the conditional Wiener  measure,
gives rise to the Feynman-Kac kernel (23) of the \it
non-selfadjoint \rm
semigroup  (suitable integrability conditions need
to be respected here as well, \cite{abc}), with the
generator $H_{\vec{A}}=
 - \nu  [\nabla  - {1\over {2\nu }}
\vec{A}(\vec{x})]^2 + c(\vec{x},t)$
defined by the right-hand-side of (33):
$${\partial _t\theta (\vec{x},t) = H_{\vec{A}}\theta (\vec{x},t)=}$$
$${
= [- \nu \triangle + \vec{A}(\vec{x}) \cdot \nabla +
{1\over 2} (\nabla
\cdot \vec{A}(\vec{x})) -
{1\over {4\nu }}[\vec{A}(\vec{x})]^2 +
c(\vec{x},t)]\theta (\vec{x},t)}\eqno (35)$$
$$= -\nu \triangle \theta (\vec{x},t)  +
\vec{A}(\vec{x})\cdot \nabla \theta (\vec{x},t) +
c_{\vec{A}}(\vec{x},t)\theta (\vec{x},t)
\enspace .$$
Here:
$${c_{A}(\vec{x},t)= c(\vec{x},t) +  {1\over 2}(\nabla
\vec{A})(\vec{x}) - {1\over {4\nu }} [\vec{A}(\vec{x})]^2\enspace .}
\eqno (36)$$
An adjoint parabolic partner of (35) reads:
$${\partial _t\theta _* = - H^*_{\vec{A}}
\theta _*  =
\nu \triangle  \theta _* +
\nabla [\vec{A}(\vec{x})\theta _*] -
c_{A}(\vec{x},t)\theta _* = }\eqno (37)  $$
$$= \nu [\nabla + {1\over {2\nu }}\vec{A}(\vec{x})]^2
\theta _*
- c(\vec{x},t)\theta _*\enspace .$$

Consequently, our assumptions (32), (33) involve  a
generalization of the adjoint parabolic system (18) to a new
adjoint one comprising  (33), (37). Obviously, the original form
of (18) is immediately restored by setting
$\vec{A}=\vec{0}$, and executing obvious replacements
$\theta _* \rightarrow u$, $\theta \rightarrow v$.

Let us emphasize again, that in contrast to Ref. \cite{abc}, where
the non-Hermitean generator $2\nu H_{\vec{A}}$, (33), has been
introduced as "the Euclidean version of the Hamiltonian"
$H=-2\nu ^2(\nabla -{i\over {2\nu }}\vec{A})^2 + \Omega $, our
electromagnetic fields stand for solutions of the usual Maxwell
equations and \it are not \rm Euclidean at all.

As long as the coefficient functions
(both additive and multiplicative) of the adjoint
parabolic system (35), (37) are not  specified, we remain
within a  general theory of positive solutions  for parabolic
equations with unbounded coefficients (of particular importance,
if we do not impose any asymptotic fall off restrictions),
\cite{krzyz,kal,bes,aron}.   The fundamental solutions, if their
existence can be  granted, usually live on  space-time strips,
and generally do not admit  unbounded  time intervals.
We shall disregard these issues at the moment, and assume the
existence of fundamental solutions without any reservations.

By exploiting  the rules of functional (Malliavin, variational)
calculus, under an assumption that we deal with
 a diffusion (in fact, Bernstein) process associated with an adjoint
pair (35), (36), it has been shown in Ref. \cite{abc} that \it
if \rm  the forward
conditonal derivatives of the process exist, then
$(D_+\vec{X})(t)= 2\nu {{\nabla \theta }\over \theta } - \vec{A} =
\vec{b}(\vec{x},t)$, (32)   and:
$${(D^2_+\vec{X})(t) = (D_+\vec{X})(t)\times curl\, \vec{A}(\vec{x})
 + \nabla \Omega (\vec{x},t)  + \nu curl\, (curl\,
 \vec{A}(\vec{x}))}\eqno (38)$$
 where $\vec{X}(0)=0$, $\vec{X}(t)=\vec{x}$, $\times $ denotes
 the vector product in $R^3$ and $2\nu c=\Omega $.

 Since $\vec{B}= curl\, \vec{A}=
 \mu _0\vec{H}$, we identify in the above
 the standard Maxwell equation  for $curl \, \vec{H}$ comprising
magnetic  effects of electric  currents in the system:
 $curl \, \vec{B} = \mu _0 [\dot{\vec{D}} + \sigma _0\vec{E}+
 \vec{J}_{ext}]$ where
 $\vec{D}=\epsilon _0\vec{E}$ while  $\vec{J}_{ext}$ represents
 external electric currents. In case of $\vec{E}=\vec{0}$,
 the external currents  only would  be relevant.
 A demand $curl\, curl\, \vec{A}=
 \nabla (\nabla \vec{A}) - \triangle \vec{A}=0$ corresponds to a
 total absence of such currents, and the Coulomb gauge choice
 $\nabla \vec{A} =0$ would leave us with harmonic functions
 $\vec{A}(\vec{x})$.

Consequently, a correct expression for the magnetically implemented
Lorentz force has appeared
on the right-hand-side of the forward acceleration formula (38),
with the forward drift (32) replacing the classical particle
velocity $\dot{\vec{q}}$ of the classical formula (25).

The above discussion implicitly involves quite sophisticated
mathematics, hence it is instructive to see that
we can bypass  the apparent complications  by directly invoking
the universal definitions (7) and (11) of conditional expectation
values, that are based on exploitation of the It\^{o}
formula only. Obviously, under an assumption that
the  Markovian diffusion process   with well defined
transition probability densities $p(\vec{y},s,\vec{x},t)$
and $p_*(\vec{y},s,\vec{x},t)$, does exist.

We shall utilize an  obvious generalization of canonical
definitions (19) of both forward and backward drifts of the
diffusion process defined by the adjoint parabolic
pair (18), as suggested by (32) with $\vec{A}=\vec{A}(\vec{x})$:
$${\vec{b}=2\nu {{\nabla \theta }\over \theta } - \vec{A}
\, \, \, , \, \, \,
\vec{b}_* = - 2\nu {{\nabla \theta _*}\over  {\theta _*}}
- \vec{A} \enspace .}\eqno (39)$$
We also demand that the corresponding adjoint equations (35), (37)
\it are \rm  solved by $\theta $ and $\theta _*$ respectively.

Taking for granted that identities $(D_+\vec{X})(t) = \vec{b}
(\vec{x},t),\, \vec{X}(t)=\vec{x}$ and $(D_-\vec{X})(t)=
\vec{b}_*(\vec{x},t)$ hold true, we can easily evaluate the
forward and backward accelerations (substitute (39), and exploit
the equations (35), (37)):
$${(D_+\vec{b})(\vec{X}(t),t)=\partial _t\vec{b} +
(\vec{b}\cdot \nabla )\vec{b} + \nu \triangle \vec{b} =}\eqno (40)$$
$$= \vec{b}\times \vec{B}  + \nu \, curl\, \vec{B}
+ \nabla \, \Omega $$
and
$${(D_-\vec{b}_*)(\vec{X}(t),t)= \partial _t \vec{b}_* +
(\vec{b}_*\cdot \nabla
)\vec{b}_* - \nu \triangle \vec{b}_*  =}\eqno (41)$$
$$= \vec{b}_*\times \vec{B} -\nu \, curl\, \vec{B}
+ \nabla \, \Omega \enspace .$$

 Let us notice that the forward and backward
acceleration formulas \it  do not \rm coincide as was the case
before ( cf. Eq. (8), (12)). There is a definite time-asymmetry
in the local description of the diffusion process in the
presence  of general magnetic fields, unless  $curl \, \vec{B} =0$.
  The quantity which is  explicitly
time-reversal invariant can be easily introduced:
$${\vec{v}(\vec{x},t)={1\over 2}(\vec{b} +
\vec{b}_*)(\vec{x},t)\Rightarrow }\eqno (42)$$
$${1\over 2}(D^2_+ + D^2_-)(\vec{X}(t))=
\vec{v}\times \vec{B} + \nabla \, \Omega \enspace .$$
As yet there is no  trace of Lorentzian electric forces, unless
extracted from the term $\nabla \, \Omega (\vec{x},t)$.  This step
we shall accomplish in Section 4.

For a probability density $\theta _* \theta =\rho $ of the
related Markovian diffusion process, \cite{zambr,olk2},
we would have fulfilled
both the  Fokker-Planck and the continuity equations:
$\partial _t\rho  = \nu \triangle \rho -
\nabla (\vec{b}\rho )= - \nabla (\vec{v}\rho )=
-\nu \triangle \rho - \nabla (\vec{b}_*\rho )$, as before
(cf. Section 1).

In the above, the equation (41) can be regarded as the Burgers
equation with a general external magnetic (plus other
external force contributions if necessary) forcing,
and its definition is an
outcome of the underlying  mathematical structure related to
the adjoint pair (33), (37) of parabolic equations.\\

Our construction shows that  solutions  of the magnetically
forced (hence nongradient) solution of the Burgers equation (41)
are  given  in the form  (39).
In reverse, the mere assumption about the decomposition of drifts,
(39), into the gradient   and nongradient part, implies that
the corresponding evolution equation (41)  is the Burgers equation
with   the nonconservative forcing. The force term has a specific
Lorentz form. Although we invoke electromagnetism, the decomposition
(39)
can be regarded to refer to an abstract nongradient  component.
In analogy to the previous Onsager-Machlup example, (24)-(28), the
fictituous Lorentz force term would arise anyway.

\section{Schr\"{o}dinger's interpolation in a constant magnetic
field and quantally inspired generalisations}

Presently, we shall confine our attention to the simplest case
of a constant   magnetic field, defined by the vector potential
$\vec{A}=\{ -{B\over 2}x_2, +{B\over 2}x_1, 0  \}$.
Here,  $\vec{B}=
\{ 0, 0, B\}$, $\nabla \vec{A}= 0$ and  $ curl\, \vec{B}= \vec{0}$
which significantly  simplifies formulas (32)-(42).

As emphasized before, most of our discussion was based on the
existence assumption for fundamental solutions of the
(adjoint) parabolic equations (33), (37).
 For magnetic fields, which do
not vanish at spatial infinities (hence for our "simplest"
choice), the situation becomes
rather complicated.  Namely, an expression for
$${c_{\vec{A}}(\vec{x},t)= c(\vec{x},t) - {{B^2}\over {16\nu }}
(x^2_1+x^2_2)}\eqno (43)$$
includes a \it repulsive \rm harmonic oscillator contribution.\\

For the existence of a well defined Markovian diffusion process
it appears necessary that a nonvanishing contribution from  an
unbounded from above $c(\vec{x},t)$  would counterbalance the
harmonic repulsion.  \\
To see that this \it must be \rm the case, let us
formally constrain $\theta (\vec{x},t)=exp(\Phi (\vec{x},t))$
to yield (in accordance with (9)) the identity:
$${c(\vec{x},t) = \partial _t\Phi  +
\nu  [\nabla \Phi ]^2 +
\nu \triangle \Phi = 0\enspace .}\eqno (44)$$
Then, we deal with the simplest version of the adjoint system
(35), (37) where, in view of $\nabla \cdot \vec{A}=0=c$, there holds:
$${\partial _t\theta =
-\nu [\nabla - {1\over {2\nu }}\vec{A}]^2\theta =
-\nu \triangle \theta + \vec{A}\cdot \nabla \theta -
{1\over {4\nu }}[\vec{A}]^2\theta }
\eqno (45)$$
$$\partial _t\theta _*=\nu [\nabla +{1\over {2\nu }}\vec{A}]^2
\theta _*
= \nu \triangle \theta _*   + \vec{A}\cdot \nabla \theta _* +
{1\over {4\nu }}[\vec{A}]^2\theta _* \enspace .$$

With  our choice, $curl \, \vec{A}=\{ 0, 0, B\} $, the
equations (45) \it do not  \rm possess a fundamental solution,
which would be well defined for \it all \rm
$(\vec{x},t)\in R^3\times R^+$:
 everything because of the harmonic repulsion term in the
 forward parabolic equation.
 We can prove (this purely mathematical argument is not
 reproduced in the present paper) that
 the function $k(\vec{y},s,\vec{x},t)$, :
$${k(\vec{y},s,\vec{x},t)=
[{B\over {4\pi sin({1\over 2}B(t-s))}}]
({1\over {2\pi (t-s)}})^{1/2}}
\eqno (46)$$
$$\times exp\{ - {1\over {2(t-s)}}(x_3-y_3)^2 - {B\over 4}
cot({B\over 2}(t-s)) [(x_2-y_2)^2+ (x_1-y_1)^2] -
 {B\over 2}(x_1y_2-x_2y_1)\} $$
\it only \rm when restricted to
 times $t-s\leq {\pi \over B}$  is an acceptable example of a
  \it unique \rm positive (actually, positivity extends to
times $t-s\leq {{ 2\pi }\over B}$) fundamental solution of
the system (44),  (rescaled to yield
$\nu \rightarrow {1\over 2}$).
Here, formally, (46) can be obtained from the expression (29)
by the  replacement $\vec{A} \rightarrow -i\vec{A}$.

An immediate insight into a harmonic repulsion  obstacle
 can be achieved after an $x-y$-plane rotation
 of Cartesian coordinates:
$x_1'=x_1cos(\omega t) - x_2sin(\omega t), \,
x_2'=x_1sin(\omega t)+x_2cos(\omega t), \, x_3'=x_3,\, t'=t$ with
$\omega ={B\over {4\sqrt{\nu }}}$. Then, equations (45) get
transformed into  an adjoint pair:
$${\partial_{t'} \theta = -\nu \triangle '\theta -
\omega ^2(x_1'^2+x_2'^2)\theta }\eqno (47)$$
$$\partial _{t'}\theta _*=\nu \triangle '\theta _*  +
\omega ^2(x_1'^2+x_2'^2)\theta _*\enspace .$$
Notice that the transformation $\omega \rightarrow i\omega$
would replace  repulsion in (47) by the harmonic attraction.
On the other hand, we can get rid of the repulsive term
by assuming that $c(\vec{x},t)$, (43) does not identically vanish.
For example, we can formally demand that, instead of (44),
$c(\vec{x},t)= +{B^2\over {8\nu }}(x_1^2+x_2^2)$ plays the
r\^{o}le of  an electric potential.
Then,  harmonic attraction  replaces repulsion
 in the final form of equations (35), (37).

As a byproduct,  we are given a  transition probability
density of the diffusion process governed by the adjoint system
(cf. (28)):
$${\partial _t\theta =
-\nu \triangle \theta + \vec{A}\cdot \nabla \theta }
\eqno (48)$$
$$\partial _t\theta _*=
 \nu \triangle \theta _*   + \vec{A}\cdot \nabla \theta _*
 \enspace .$$
with $\vec{A}={B\over 2}\{ -x_2, x_1, 0 \}$.
Namely, by means of the previous $x-y$-plane rotation, (48)
is transformed into a pair of time adjoint heat equations:
$${\partial _{t'}\theta = - \nu \triangle '\theta \, \, \, ,\, \, \,
\partial _{t'}\theta _* = \nu \triangle '\theta _* }\eqno (49)$$
whose fundamental solution is the standard heat kernel.

Finding explicit analytic solutions of rather involved
equations (35), (37) is a formidable task on its own,
in contrast to much simpler-unforced or conservatively forced
dynamics issue.

Interestingly, we can produce a number of examples  by invoking
the quantum Schr\"{o}dinger dynamics.   This
quantum inspiration has been proved to be very useful in the past,
\cite{zambr,zambr1}.
At this point, we shall follow the  idea of Ref. \cite{olk2}
where the strategy developed for solving the Schr\"{o}dinger boundary
data problem has been applied to quantally induced stochastic
processes (e.g.  Nelson's diffusions, \cite{nel,nel1}).
They were considered  as  a particular case  of the
general theory appropriate for nonequilibrium statistical
physics processes as governed by the adjoint pair (18), and
exclusively in conjunction with Born's statistical postulate
in quantum theory.

The Schr\"{o}dinger picture  quantum evolution  is then
consistently representable as a Markovian diffusion process. All
that follows from the previously outlined Feynman-Kac kernel route,
\cite{nel,nel1,zambr,olk2,blanch,olk,olk1}, based on exploiting
the adjoint pairs of parabolic equations.
However, the respective semigroup theory has been developed for
pure gradient drift fields, hence without reference to
any impact of electromagnetism on the pertinent diffusion process.
And electromagnetism is definitely ubiquitous in the world of
quantum phenomena.

Let us start from an ordinary Schr\"{o}dinger equation
 for a charged particle in  an arbitrary
external electromagnetic field, in its standard dimensional form.
To conform with the previous notation let us absorb the charge
$e$ and mass $m$ parameters in the definition of $\vec{A}(\vec{x})$
and the potential $\phi (\vec{x})$, e.g. we consider
$B$ instead of ${e\over m}B$ and $\phi $ instead $\phi /m$.
Additionally, we set $\nu $ instead of ${\hbar \over {2m}}$.
Then, we have:
$${i\partial _t \psi (\vec{x},t)= - \nu (\nabla  -
{i\over {2\nu }}\vec{A})^2 \psi (\vec{x},t) + {1\over {2\nu }}
\phi (\vec{x})\psi (\vec{x},t)\enspace . }\eqno (50)$$

The standard Madelung substitution   $\psi = exp(R+iS)$ allows
to introduce the real functions $\theta = exp(R+S)$ and
$\theta _*=exp(R-S)$ instead of complex ones
$\psi, \overline {\psi }$.
They are solutions of an adjoint  parabolic system (35), (37),
where the impact of (50) is encoded in a specific functional form
of, otherwise arbitrary potential $c(\vec{x},t)$:
$${c(\vec{x},t)= {1\over {2\nu }} \Omega (\vec{x},t)=
{1\over {2\nu }} [2Q(\vec{x},t)-\phi (\vec{x})]}
\eqno (51)$$
$$Q(\vec{x},t)=2\nu ^2{{\triangle \rho ^{1/2}(\vec{x},t)}\over {\rho
^{1/2}(\vec{x},t)}}=2\nu ^2 \bigl [ \triangle R(\vec{x},t) + [\nabla
R(\vec{x},t)]^2 \bigr ]\enspace . $$

The quantum probability density $\rho (\vec{x},t)=
\psi (\vec{x},t)\overline {\psi }(\vec{x},t)=
\theta (\vec{x},t)\theta _*(\vec{x},t)$  displays a factorisation
 $\rho =\theta \theta _*$ in terms of solutions of adjoint parabolic
 equations, which we recognize to be characteristic for probabilistic
 solutions  (Markov diffusion processes)
 of the Schr\"{o}dinger boundary data problem (cf. Section 1),
 \cite{zambr,blanch,olk2,olk}.
It is  easy to verify the validity of the Fokker-Planck equation
whose forward drift has the  form (38).  Also, equations (40), (41)
do follow with $\Omega =2Q - \phi $.

By defining $\vec{E}=-\nabla \phi $
(with $\phi $ utilised instead of
${e\over m}\phi $), we immediately arrive at the complete Lorentz
force contribution in all acceleration formulas
(before, we have used $curl \, \vec{B}=0$):
$${\partial _t\vec{b} + (\vec{b}\cdot \nabla )\vec{b} +
\nu \triangle \vec{b} = \vec{b}\times \vec{B} + \vec{E} +
\nu curl \, \vec{B} +  2\nabla Q
}\eqno (52)$$
$$\partial _t\vec{b}_* + (\vec{b}_*\cdot \nabla )\vec{b}_* -
\nu \triangle
\vec{b}_* = \vec{b}_* \times \vec{B} + \vec{E} - \nu curl \, \vec{B}
+ 2\nabla Q\enspace .$$
Moreover, the velocity field named the current velocity of the
flow, $\vec{v}={1\over 2}(\vec{b} + \vec{b}_*)$,
enters  the familiar local conservation laws
(see also \cite{blanch} for a discussion of how the "quantum potential"
$Q$ affects such  laws in case of the standard Brownian
motion and Smoluchowski-type diffusion processes)
$${\partial \rho = - \nabla (\vec{v}\rho )}\eqno (53)$$
$$\partial _t\vec{v} +(\vec{v}\cdot \nabla )\vec{v} =
\vec{v}\times \vec{B} + \vec{E}  + \nabla Q \enspace .$$

A comparison with (33)-(43) shows that equations (50)-(53) can
be regarded  as  the  specialized version  of the general external
forcing problem with an explicit electromagnetic (Lorentz
force-inducing) contribution and an
arbitrary  term  of non-electromagnetic origin, which we denote by
$c(\vec{x},t)$ again. Obviously,  $c$ is represented in (51),
 by ${1\over {\nu }}Q(\vec{x},t)$.

We have therefore arrived at the
following ultimate generalization of the adjoint parabolic
system (18), that encompasses the nonequilibrium statistical
physics and essentially quantum evolutions on an equal footing
(with no clear-cut discrimination between these options, as in
Ref. \cite{olk2})
and gives rise to an external (Lorentz)  electromagnetic forcing:
$${\partial _t\theta (\vec{x},t)=[ - \nu (\nabla  -
{1\over {2\nu }}\vec{A})^2 -
{1\over {2\nu }}\phi (\vec{x}) + c(\vec{x},t)] \theta
(\vec{x},t)}\eqno (54)$$
$$\partial _t\theta _*(\vec{x},t)= [\nu (\nabla +
 {1\over {2\nu }}\vec{A})^2
 + {1\over {2\nu }}\phi (\vec{x}) -
c(\vec{x},t)]\theta _*(\vec{x},t)\enspace . $$
A subsequent  generalisation encompassing time-dependent
electromagnetic  fields is immediate.

The adjoint parabolic pair  (54) of equations can  thus be
regarded to determine a Markovian diffusion
process in exactly the same way as (18) did.
If only a suitable choice of  vector and scalar potentials in (54)
guarrantees  a continuity and positivity of the involved semigroup
kernel (take the Radon-Nikodym density of the form (34), with
$\Omega  \rightarrow -\phi + \Omega $ , and integrate with
respect to the  conditional Wiener measure), then the mere
knowledge of such integral  kernel
suffices  for the implementation of steps (18)-(22), with
$u\rightarrow \theta _*$, $v\rightarrow \theta $.  To this end
it is not at all necessary that $k(\vec{x},s,\vec{y},t)$ is a
fundamental solution  of (54). A sufficient condition  is that
the semigroup kernel is a
continuous (and positive) function. The kernel  may not even be
differentiable, see e.g. Ref. \cite{olk2} for a discussion of
that issue  which is typical for quantal situations.

After adopting (54) as the principal dynamical ingredient of
the electromagnetically forced Schr\"{o}dinger interpolation,
we  must slightly adjust  the emerging acceleration
formulas. Namely, they have the form (52) , but we
need to  replace $2Q(\vec{x},t)$  by, from now on
 arbitrary,  potential $\Omega (\vec{x},t)= 2\nu c(\vec{x},t)$.
The  second equation in (53) also takes a new  form:
$${\partial _t\vec{v} + (\vec{v}\cdot \nabla )\vec{v} =
\vec{v} \times \vec{B} + \vec{E} + \nabla (\Omega  - Q)}
\eqno (55)$$
see e.g. Ref. \cite{blanch} for more detailed explanation of
this step.  The presence in (54) of the  density-dependent
$- \nabla Q$ term  finds its origin  in the  identity
$\vec{b}-\vec{b}_* = 2\nu \nabla \rho (\vec{x},t)$ and is a necessary
consequence of the involved (forced in the present case)  Brownian
motion, see e.g. \cite{gei,garb1,vig}.

Finally, the second equation  (52) with $\Omega $ replacing $2Q$
is the most general form of the Burgers equation with  an
external forcing, where  the electromagnetic (Lorentz force)
contribution has been extracted for convenience. Solutions of
this equation must be sought for  in the form (39), which
 generalizes the logarithmic Hopf-Cole transformation to
 non-gradient drift  fields.
 Equations (54) are the associated
 parabolic partial differential (generalised heat) equations, that
 completely determine probabilistic solutions (Markovian diffusion
 processes) of the  Schr\"{o}dinger boundary data (interpolation)
 problem. In turn, for this particular random transport,
  the forced Burgers velocity fields play the r\^{o}le of  backward
  drifts of the process.

\section{Outlook}

Our discussion, albeit motivated by the issue of diffusive
 matter transport that is consistently driven by Burgers velocity
 fields  (this extends both to  the compressible and
 incompressible cases), has little to do with classical fluids.
 The emergence of shock pressure fronts is more natural in the
 compressible situation.   This shock profile possibility (inherent
 to the Burgers equation) has been imported to the nonequilibrium
 statistical physics of random phenomena by exploring  the idea of
 Schr\"{o}dinger's  interpolation problem and revealing its
  connection with the Burgers dynamics. That has been the subject
  of Section 1.\\
  The next important result (a preliminary discussion of rotational
  Burgers fields can be found in Ref. \cite{sai}) amounts to relaxing
  the gradient-field assumption (that is crucial for the validity of the
  Hopf-Cole transformation). In Sections 2 and 3 we have analyzed the
  ways to generalise the Feynman-Kac kernel strategy  so that the
  involved
  (drifts) velocity fields admit the nongradient form.
  Our analysis was perfomed with rather explicit electromagnetic
 connotations.  Equations (35) and (37) generalise the adjoint pair (18)
 to diffusion processes with nongradient drifts, (39). \\
 As follows from (41), the very presence of the nongradient term  in
 the decomposition (39)
 implies that   the corresponding evolution equation for the velocity
 field (backward drift of the process) is the Burgers equation with the
 specific Lorentz-type forcing.\\
 Section 4 extends the discussion to quantally implemented diffusion
 processes, where the minimal electromagnetic coupling is a celebrated
 recipe.  This quantal motivation allows to arrive at the adjoint
 system (54), that incorporates an electric contribution and allows
 to define and solve the Burgers equation with the combined
 conservative and  nonconservative
 (electromagnetic, in particular) forcing.
 Let us emphasise again that a transformation of the Burgers
 equation (whatever the force term is) into a generalised diffusion
 equation is not merely a formal linearisation trick.
 This, \cite{burg},   "nonlinear diffusion equation" does indeed
 refer to a well defined stochastic diffusion process,
 but  a complete information about its
 features is encoded in the involved parabolic equations.

 \vskip0.5cm
{\bf Acknowledgement}: Two of the  authors (P. G. and R. O. )
received  a financial support from the KBN research grant
No 2 P302 057 07.  P. G. would like to express his gratitude to
Professors Ana Bela Cruzeiro and Jean-Claude Zambrini for
enlightening discussions.


\begin{thebibliography}{99}


\bibitem{burg} J. M. Burgers, "The Nonlinear Diffusion Equation",
Reidel,  Dordrecht, 1974

\bibitem{hopf} E. Hopf, Commun.  Pure Appl. Math., 3, (1950), 201

\bibitem{zeld} S. F. Shandarin, B. Z. Zeldovich, Rev. Mod. Phys. 61,
(1989),
185

\bibitem{alb} S. Albeverio, A. A. Molchanov, D. Surgailis, Prob.
Theory
Relat. Fields,

100, (1994), 457

\bibitem{woy}  Y. Hu, W. A. Woyczynski, pp. 135,
in: "Chaos-The Interplay 
Between

Stochastic  and  Deterministic Behaviour", eds. P. Garbaczewski,
M. Wolf,

A. Weron, LNP vol. 457, Springer-Verlag, Berlin, 1995

\bibitem{gurb} S. N. Gurbatov, A. I. Saichev, Soviet JETP
(Russian Edition), 
80, (1981), 689 

\bibitem{frisch} Z. She, E. Aurell, U. Frisch, Commun. Math. Phys.
148, 
(1992), 623

\bibitem{sinai} Ya. G. Sinai, Commun. Math. Phys., 148, (1992), 601

\bibitem{fournier} J.D. Fournier, U. Frisch, J. Mec. Theor. Appl., 2, 
(1983), 699

\bibitem{woy1} W. A. Woyczynski, pp. 279, in: "Nonlinear Waves and
Weak 
Turbulence", 

eds. N. Fitzmaurice et al, Birka\"{a}user, Boston, 1993

\bibitem{parisi} J. P. Bouchaud, M. M\'{e}zard, G. Parisi, Phys. Rev. 
E 52, (1995), 3656

\bibitem{pol} A. M. Polyakov, Phys. Rev. E 52, (1995), 6183

\bibitem{parisi1} M. Kardar, G. Parisi, Y. C. Zhang, Phys. Rev. Lett. 
56, (1986), 889

\bibitem{siggia} B. I. Shraiman, E. D. Siggia, Phys. Rev. E 49,
(1994), 
2912

\bibitem{yak} A. Chekhlov, V. Yakhot, Phys. Rev. E 51, (1995), 2739

\bibitem{monin} A. S. Monin, A. M. Yaglom, "Statistical Fluid
Mechanics", 
The MIT Press, 

Cambridge, Mass. 1973

\bibitem{gurba} S. N. Gurbatov, A. N. Malakhov, A. I. Saichev,
"Nonlinear Random

Waves and Turbulence in Nondispersive Media: Waves, Rays,

Particles", Manchester University Press, Manchester, 1991

\bibitem{truman} A. Truman,  H. Z. Zhao, J. Math. Phys. 37, (1996),
283

\bibitem{holden} H. Holden, T. Lindstr{\o}m, B. {\O}ksendal, J.
Ub{\o}e, 
T. S. Zhang, 

Commun. Part. Diff. Eq., 19, (1994), 119

\bibitem{walsh} J. B. Walsh, pp. 265, in: \'{E}cole d'\'{E}t\'{e}
de 
Probabilit\'{e}s  de Saint-Flour XIV,

eds. R. Carmona, H. Kesten, J. B. Walsh, LNM vol. 1180,
Springer-Verlag,

Berlin, 1986

\bibitem{saf} P. G. Saffman, J. Fluid Mech. 8, (1960), 273

\bibitem{town} A. A. Townsend, Proc. Roy.Soc. A 209, (1951), 418

\bibitem{sai} A. I. Saichev, W. A. Woyczynski, Physica D, in press


\bibitem{woy2} A. I. Saichev, W. A. Woyczynski, SIAM J. Appl. Math.
in press

\bibitem{fried} A. Friedman, "Partial Differential Equations of
Parabolic type", Prentice-Hall,

Englewood, NJ, 1964

\bibitem{horst} W. Horsthemke, R. Lefever, "Noise-Induced
Transitions",
Springer-Verlag,

Berlin, 1984


\bibitem{kean} H. P. Mc Kean, pp. 177,  in: "Lecture Series in
Differential
Equations", vol. II,

ed. A. K. Aziz, Van Nostrand, Amsterdam, 1969

\bibitem{cald} P. Calderoni, M. Pulvirenti, Ann. Inst. Henri
Poincar\'{e},
39, (1983), 85

\bibitem{osa} H. Osada, S. Kotani, J. Math. Soc. Japan, 37,
(1985), 275


\bibitem{krzyz} M. Krzy\.{z}a\'{n}ski, A. Szybiak,
Lincei-Rend. Sc.
fis. mat. e nat. 28, (1959), 26


\bibitem{olk2} P. Garbaczewski, R. Olkiewicz, J. Math. Phys. 37,
(1996), 730

\bibitem{blanch} Ph. Blanchard, P.Garbaczewski, Phys. Rev. E 49,
(1994), 3815


\bibitem{haus} U. G. Haussmann, E. Pardoux, Ann. Prob. 14, (1986),
1188

\bibitem{fol}   H. F\"{o}llmer, pp. 119, in: "Stochastic
Processes-Mathematics and

Physics", eds. S. Albeverio, Ph. Blanchard, L. Streit, LNP vol. 1158,

Springer-Verlag, Berlin, 1985

\bibitem{nel} E. Nelson, "Quantum Fluctuations", Princeton University
Press, Princeton, 1985


\bibitem{zambr} J. C. Zambrini, J. Math. Phys. 27, (1986), 3207

\bibitem{zambr1} J. C. Zambrini, pp. 393, in: "Chaos-The Interplay
Between Stochastic

and Deterministic Behaviour", eds. P. Garbaczewski, M. Wolf, A. Weron,

LNP vol 457, Springer-Verlag, Berlin, 1995

\bibitem{nel1} E. Nelson, "Dynamical Theories of the Brownian Motion",
Prinecton

University Press, Princeton, 1967



\bibitem{vig} P. Garbaczewski, J. P. Vigier, Phys. Rev. A46, (1992),
4634

\bibitem{olk} P.Garbaczewski, R. Olkiewicz, Phys. Rev. A 51, (1995),
3445


\bibitem{olk1} P.Garbaczewski, J. R. Klauder, R. Olkiewicz,
Phys. Rev. E 51, (1995), 4114



\bibitem{flem} W. H. Fleming, H. M. Soner, "Controlled  Markov
Processes
and  Viscosity

Solutions", Springer-Verlag, Berlin, 1993

\bibitem{alb1} S. Albeverio, Ph. Blanchard, R. H{\o}egh-Krohn, pp.
189,
in: "Stochastic

Aspects of Classical and Quantum Systems", eds. S. Albeverio et al,

LNM vol. 1109, Springer-Verlag, Berlin, 1985

\bibitem{freid} M. Freidlin, "Functional Integration and Partial
Differential Equations", 

Princeton University press, Princeton, 1985


\bibitem{flem} W. H. Fleming, H. M. Soner, "Controlled  Markov
Processes
and  Viscosity

Solutions", Springer-Verlag, Berlin, 1993

\bibitem{garb} P. Garbaczewski, Phys. Lett. A 175, (1993), 7

\bibitem{alb1} S. Albeverio, Ph. Blanchard, R. H{\o}egh-Krohn, pp.
189,
in: "Stochastic

Aspects of Classical and Quantum Systems", eds. S. Albeverio et al,

LNM vol. 1109, Springer-Verlag, Berlin, 1985




\bibitem{simon} B. Simon, "Functional Integration and Quantum
Physics", Academic Press,

New York, 1979

\bibitem{glimm} J. Glimm, A. Jaffe, "Quantum Physics-A Functional 
Integral Point 

of View", Springer-Verlag, Berlin, 1981

\bibitem{schuss} Z. Schuss, "Theory and Applications of Stochastic 
Differential

Equations", Wiley, NY, 1980

\bibitem{hub} J. B. Hubbard, P. G. Wolynes, J. Chem. Phys. 75, (1981), 
3051

\bibitem{sung} W. Sung, H. L. Friedman, J. Chem. Phys. 87, (1987), 649

\bibitem{mori} H. Mori, Progr. Theor. Phys. 33, (1965), 243

\bibitem{izm} A. F. Izmailov, S. Arnold, S. Holler, A. S. Myerson,
Phys. Rev. E 52,

(1995), 1325


\bibitem{has1} H. Hasegawa, Progr. Theor. Phys. 56, (1976), 44

\bibitem{hunt} K. L. C. Hunt, J. Chem. Phys. 75, (1981), 976

\bibitem{ronc} M. Roncadelli, Phys. Rev. E 52, (1995), 4661

\bibitem{wieg} F. W. Wiegel, J. Ross, Phys. Lett. 84 A, (1981), 465

\bibitem{roep} G. Roepstorff, "Path Integral Approach to Quantum
Physics", Springer-Verlag,

Berlin, 1994

\bibitem{simon1} J. Avron, I. Herbst, B. Simon, Duke Math. Journ. 45,
(1978), 847

\bibitem{simon2}  J. Avron, I. Herbst, B. Simon,   Commun. Math. Phys.
79, (1981), 529


\bibitem{kal} A. M. Ilin, A. S. Kalashnikov, O. A. Oleinik, Usp. Mat.
Nauk  (in Russian),

27, (1962), 65 

\bibitem{bes} P. Besala, Ann. Polon. Math. 29, (1975), 403

\bibitem{aron} D. G. Aronson, P. Besala, Colloq. Math. 18, (1967),
125

\bibitem{abc} A. B. Cruzeiro, J. C. Zambrini, J. Funct. Anal. 96,
(1991), 62

\bibitem{gei} B. T. Geilikman, Sov. JETP, (Russian Edition), 17,
(1947),   830

\bibitem{garb1} P. Garbaczewski, Phys. Lett. A 162, (1992), 129

\end{thebibliography}
\end{document}